\documentclass[preprint,,trackchanges]{aastex63}

\newcommand{\mgii}{Mg\thinspace\textsc{ii}\ }

\newcommand{\lr}[1]{{\color{black}{#1}}{}}

\usepackage{enumitem}
\usepackage{graphicx}

\received{May 31, 2022}
\revised{May 31, 2022}

\submitjournal{ApJ}

\shorttitle{CLASP2 Quiet Sun Center-to-limb Variation}
\shortauthors{Rachmeler et al.}
\setwatermarkfontsize{80 pt}

\begin{document}


\title{Quiet Sun Center to Limb Variation of the Linear Polarization Observed by \\CLASP2 Across the Mg II h \& k Lines}

\correspondingauthor{L.~A. Rachmeler}
\email{laurel.rachmeler@noaa.gov}

\author[0000-0002-3770-009X]{L.~A. Rachmeler}
\affiliation{NOAA National Centers for Environmental Information, 325 Broadway, Boulder, CO 80305, USA}

\author[0000-0001-5131-4139]{J.~Trujillo Bueno}
\affiliation{Instituto de Astrof\'{i}sica de Canarias, E-38205 La Laguna, Tenerife, Spain}
\affiliation{Departamento de Astrof\'{i}sica, Universidad de La Laguna, E-38206 La Laguna, Tenerife, Spain}
\affiliation{Consejo Superior de Investigaciones Cient\'{i}ficas, Spain}

\author[0000-0002-9921-7757]{D.~E. McKenzie}
\affiliation{NASA Marshall Space Flight Center, 320 Sparkman Dr.~NW, Mail Stop ST13, Huntsville, AL 35805, USA}

\author[0000-0001-8830-0769]{R. Ishikawa}
\affiliation{National Astronomical Observatory of Japan, National Institutes of Natural Science, 2-21-1 Osawa, Mitaka, Tokyo 181-8588, Japan}

\author[0000-0003-0972-7022]{F. Auch\`{e}re}
\affiliation{Institut d'Astrophysique Spatiale, CNRS, Univ. Paris-Sud, Universit\'e Paris-Saclay, B\^at.\ 121, 91405 Orsay, France}

\author{K. Kobayashi}
\affiliation{NASA Marshall Space Flight Center, 320 Sparkman Dr.~NW, Mail Stop ST13 Huntsville, AL 35805, USA}

\author[0000-0002-2093-085X]{R. Kano}
\affiliation{National Astronomical Observatory of Japan, National Institutes of Natural Science, 2-21-1 Osawa, Mitaka, Tokyo 181-8588, Japan}

\author[0000-0003-3765-1774]{T.~J. Okamoto}
\affiliation{National Astronomical Observatory of Japan, National Institutes of Natural Science, 2-21-1 Osawa, Mitaka, Tokyo 181-8588, Japan}

\author[0000-0001-9076-6461]{C.~W. Bethge}
\affiliation{NOAA National Centers for Environmental Information, 325 Broadway, Boulder, CO 80305, USA}
\affiliation{Cooperative Institute for Research in Environmental Sciences, 325 Broadway, Boulder, CO 80305, USA}

\author[0000-0003-3034-8406]{D. Song}
\affiliation{Korea Astronomy and Space Science Institute, 776 Daedeok-daero, Yuseong-gu, Daejeon 34055, Republic of Korea}
\affiliation{National Astronomical Observatory of Japan, National Institutes of Natural Science, 2-21-1 Osawa, Mitaka, Tokyo 181-8588, Japan}

\author[0000-0001-9095-9685]{E.~Alsina Ballester}
\affiliation{Instituto de Astrof\'{i}sica de Canarias, E-38205 La Laguna, Tenerife, Spain}
\affiliation{Universidad de La Laguna, Departamento de Astrof\'{i}sica, E-38206 La Laguna, Tenerife, Spain}

\author[0000-0002-8775-0132]{L.~Belluzzi}
\affiliation{Istituto Ricerche Solari (IRSOL), Università della Svizzera italiana (USI), CH-6605 Locarno-Monti, Switzerland}
\affiliation{Leibniz-Institut f\"{u}r Sonnenphysik (KIS), D-79104 Freiburg, Germany}
\affiliation{Euler Institute, Universit\`{a} della Svizzera italiana (USI), CH-6900 Lugano, Switzerland}

\author{T.~del Pino Alem\'{a}n}
\affiliation{Instituto de Astrof\'{i}sica de Canarias, E-38205 La Laguna, Tenerife, Spain}
\affiliation{Universidad de La Laguna, Departamento de Astrof\'{i}sica, E-38206 La Laguna, Tenerife, Spain}

\author[0000-0003-4088-7673]{M. Yoshida}
\affiliation{National Astronomical Observatory of Japan, National Institutes of Natural Science, 2-21-1 Osawa, Mitaka, Tokyo 181-8588, Japan}
\affiliation{Department of Astronomical Science, School of Physical Sciences, SOKENDAI (The Graduate University for Advanced Studies), Mitaka, Tokyo 181-8588, Japan.}

\author{T. Shimizu}
\affiliation{Institute of Space and Astronautical Science, Japan Aerospace Exploration Agency, Sagamihara, Kanagawa 252-5210, Japan.}

\author[0000-0002-5608-531X]{A. Winebarger}
\affiliation{NASA Marshall Space Flight Center, 320 Sparkman Dr.~NW, Mail Stop ST13, Huntsville, AL 35805, USA}

\author[0000-0002-4691-1729]{A.~R. Kobelski}
\affiliation{NASA Marshall Space Flight Center, 320 Sparkman Dr.~NW, Mail Stop ST13, Huntsville, AL 35805, USA}

\author[0000-0002-7219-1526]{G.~D. Vigil}
\affiliation{NASA Marshall Space Flight Center, 320 Sparkman Dr.~NW, Mail Stop ST13, Huntsville, AL 35805, USA}

\author[0000-0002-8370-952X]{B. De Pontieu}
\affiliation{Lockheed Martin Solar \& Astro-physics Laboratory, Palo Alto, CA 94304, USA}
\affiliation{Rosseland Centre for Solar Physics, University of Oslo, NO-0315 Oslo, Norway}
\affiliation{Institute of Theoretical Astrophysics, University of Oslo, NO-0315 Oslo, Norway}

\author[0000-0002-6330-3944]{N. Narukage}
\affiliation{National Astronomical Observatory of Japan, National Institutes of Natural Science, 2-21-1 Osawa, Mitaka, Tokyo 181-8588, Japan}

\author[0000-0001-5616-2808]{M. Kubo}
\affiliation{National Astronomical Observatory of Japan, National Institutes of Natural Science, 2-21-1 Osawa, Mitaka, Tokyo 181-8588, Japan}

\author[0000-0003-2991-4159]{T. Sakao}
\affiliation{Institute of Space and Astronautical Science, Japan Aerospace Exploration Agency, Sagamihara, Kanagawa 252-5210, Japan}
\affiliation{Department of Space and Astronautical Science, School of Physical Sciences, SOKENDAI (The Graduate University for Advanced Studies), Sagamihara, Kanagawa 252-5210, Japan}

\author{H. Hara}
\affiliation{National Astronomical Observatory of Japan, National Institutes of Natural Science, 2-21-1 Osawa, Mitaka, Tokyo 181-8588, Japan}

\author[0000-0003-4452-858X]{Y. Suematsu}
\affiliation{National Astronomical Observatory of Japan, National Institutes of Natural Science, 2-21-1 Osawa, Mitaka, Tokyo 181-8588, Japan}

\author[0000-0002-8292-2636]{J.~\v{S}t\v{e}p\'{a}n}
\affiliation{Astronomical Institute of the Czech Academy of Sciences, Fri\v{c}ova 298, 25165 Ond\v{r}ejov, Czech Republic}

\author[0000-0001-9218-3139]{M. Carlsson}
\affiliation{Rosseland Centre for Solar Physics, University of Oslo, NO-0315 Oslo, Norway}
\affiliation{Institute of Theoretical Astrophysics, University of Oslo, NO-0315 Oslo, Norway}

\author[0000-0003-4936-4211]{J. Leenaarts}
\affiliation{Institute for Solar Physics, Department of Astronomy, Stockholm University, AlbaNova University Centre, SE-106 91, Stockholm, Sweden}

\author{A. Asensio Ramos}
\affiliation{Instituto de Astrof\'{i}sica de Canarias, E-38205 La Laguna, Tenerife, Spain}
\affiliation{Universidad de La Laguna, Departamento de Astrof\'{i}sica, E-38206 La Laguna, Tenerife, Spain}










\begin{abstract}

The CLASP2 (Chromospheric LAyer SpectroPolarimeter 2) sounding rocket mission was launched on 2019 April 11. 
CLASP2 measured the four Stokes parameters of the \mgii $h$ \& $k$ spectral region around 2800\AA\ along a 200\arcsec\ slit at three locations on the solar disk, achieving the first spatially and spectrally resolved observations of the solar polarization in this near ultraviolet region. 
The focus of the work presented here is the center-to-limb variation of the linear polarization across these resonance lines, which is produced by the scattering of anisotropic radiation in the solar atmosphere.   
The linear polarization signals of the \mgii $h$ \& $k$ lines are sensitive to the magnetic field from the low to the upper chromosphere through the Hanle and magneto-optical effects. 
We compare the observations to theoretical predictions from radiative transfer calculations in unmagnetized semi-empirical models, arguing that  magnetic fields and horizontal inhomogeneities are needed to explain the observed polarization signals and spatial variations. 
This comparison is an important step in both validating and refining our understanding of the 
physical origin of these polarization signatures, and also in paving the way toward future space telescopes for probing the magnetic fields of the solar upper atmosphere via ultraviolet spectropolarimetry.

\end{abstract}

\keywords{Polarization, Scattering, Sun: magnetic fields, Sun: chromosphere, Sun: transition region, Sun: UV radiation} 



\section{Introduction} \label{sec:intro}

Our ability to acquire empirical information on the magnetic fields that permeate the atmospheres of the Sun and of other stars largely relies on two key components: the instrumentation needed to measure the polarization of the emitted spectral line radiation, and knowledge gained from theoretical modeling investigations which is required to interpret these spectropolarimetric observations. 
The advent of routine monitoring of solar surface magnetic fields using the polarization produced by the Zeeman effect in photospheric spectral lines catalyzed many new advances in solar physics.           
It is now a routine matter to estimate the magnetic field throughout the solar atmosphere by combining these photospheric measurements with extrapolation techniques based on simplifying assumptions of force-free fields, which is a reasonable approximation in the solar corona. 
However, the chromosphere is not a force-free region, and this leads to large errors in the extrapolated field  \citep[e.g.,][]{derosa2009}. 
Routine measurements of the magnetic field at the base of the corona---or the top of the chromosphere---would significantly improve our ability to accurately model the coronal field and provide key information on the physics of the chromosphere and transition region.
This magnetic field is the key missing measurement needed to advance our understanding of the solar atmosphere as it drives the mass and energy flow, and dominates the energy budget.

The CLASP \citep[Chromospheric Lyman-Alpha Spectro-Polarimeter,][]{Kobayashi2012,Kano2012SPIE} and CLASP2 
\citep[Chromospheric LAyer Spectro-Polarimeter 2,][]{Narukage2016,Ishikawa2021} sounding rocket missions were developed to explore the possibility of mapping the magnetic field in the upper solar chromosphere by providing the first measurements of polarization in prominent ultraviolet (UV) lines. In 2015, the CLASP instrument measured the linear polarization of the hydrogen Lyman-$\alpha$ line at $1215.7$\,\r{A} \citep{Kano2017} and of the Si {\sc iii} resonance line at $1206.5$\,\r{A} \citep{Ishikawa2017}in a quiet region of the solar disk.
In 2019, CLASP2 measured the linear and circular polarization of the \mgii $h$ \& $k$ doublet near $2800$\,\r{A} in quiet and active regions of the solar disk (Figs. \ref{fig:targets} \& \ref{fig:prettyspectra}).
The polarization observed in these lines is due to the combined action of anisotropic radiation pumping and the Hanle and Zeeman effects in the optically thick plasma of the solar chromosphere \citep[for a review, see][]{JTB2017}.

Over the last few years, a number of theoretical investigations have enhanced our understanding of the physical mechanisms responsible for the polarization across the \mgii $h$ \& $k$ lines and its magnetic sensitivity \citep{Belluzzi2012, AlsinaBallester2016, delpinoaleman2016, delpinoaleman2020}. 
The linear polarization---Stokes $Q$ \& $U$---is controlled by the combined action of scattering processes and the Hanle and magneto-optical (MO) effects, while the circular polarization---Stokes $V$---is dominated by the Zeeman effect.

In one-dimensional (1D) {\it unmagnetized} models of the solar atmosphere, 
the polarization is produced only by the scattering of anisotropic radiation, 
which produces broad $Q/I$\footnote{As usual, the reference direction for $+Q$ is parallel to the nearest solar limb.} profiles with a clear center-to-limb variation (CLV), 
and $U/I=0$ everywhere. 
The red line in Figure~\ref{fig:limb_theoryVobs1} shows this theoretical $Q/I$ line profile near the limb.
At the center of the \mgii $k$ line $Q/I{>}0$ (Fig.~\ref{fig:limb_theoryVobs1}, A), 
whereas $Q/I{=}0$ at the very center of the (intrinsically unpolarizable) $h$ line (Fig.~\ref{fig:limb_theoryVobs1}, C). 
The \mgii $k$ line near-wings have negative $Q/I$ peaks (Fig.~\ref{fig:limb_theoryVobs1}, B), 
with the blue negative peak deeper than the red one, 
while $Q/I$ shows an antisymmetric feature in the near wings around the center of the $h$ line. 
Such near-wing features are produced by the joint action 
of partial frequency redistribution (PRD) and quantum-mechanical 
interference between the substates pertaining to the two upper levels 
of the \mgii $h$ \& $k$ lines (hereafter, $J$-state interference). 
The combined action of these physical ingredients is also the cause 
of the predicted negative $Q/I$ in the wing wavelengths between the $h$ \& $k$ lines (Fig.~\ref{fig:limb_theoryVobs1}, D), 
and of the large positive $Q/I$ values expected for the far blue and red wings of the $h$ and $k$ lines (Fig.~\ref{fig:limb_theoryVobs1}, E). 

In 1D {\it magnetized} models of the solar atmosphere, the Hanle effect modifies the line-center $Q/I$ and $U/I$ signals of the \mgii $k$ line, while the $\rho_VQ$ and $\rho_VU$ magneto-optical terms of the Stokes $U$ and $Q$ transfer equations create $U/I$ wing signals and introduce magnetic sensitivity in the wings of both $Q/I$ and $U/I$ \citep{AlsinaBallester2016, delpinoaleman2016, delpinoaleman2020}. 
In contrast to 1D models, the plasma of the solar atmosphere is dynamic and inhomogeneous in both the horizontal and vertical directions, and the ensuing breaking of the axial symmetry of the incident radiation field at each point within the medium can produce changes in the $Q/I$ and $U/I$ line-center and wing signals even in the absence of a magnetic field \citep[e.g.,][]{MansoSainz2011,Stepan2016}.
          
Very few observations of the solar \mgii $h$ \& $k$ polarization exist.  
A linear polarization dataset taken at different times close to the limb at 10 wavelengths in the far wings of the \mgii $h$ \& $k$ lines was obtained by the Ultraviolet Spectrometer and Polarimeter \citep[UVSP;][]{Woodgate1980} on board the Solar Maximum Mission \citep[SMM;][]{Bohlin1980}.
This dataset was useful to test the theoretical predictions of \citet{Auer1980} for the linear polarization in the far wings of the lines, namely positive $Q/I$ signals in the far blue and red wings of the lines and negative $Q/I$ in the wing between the lines 
(these predictions were obtained using the approximation of coherent scattering in the observer's frame).
The data were initially analyzed by \cite{Henze1987}, who found the strong positive $Q/I$ signals that \citet{Auer1980} had estimated for the far blue and red wings of the $h$ \& $k$ lines, respectively. 
Recently, a re-calibration of the same limb data has detected the expected negative $Q/I$ signals between the $h$ \& $k$ lines \citep{MansoSainz2019}, the true physical origin of which is the joint action of PRD effects and $J$-state interference \citep{Belluzzi2012}. 
Before CLASP2, this UVSP data was the only existing observation of the \mgii far wings polarization, albeit at very low spectral and spatial resolution. 
The CLASP2 data are the first spectrally and spatially resolved polarization observations of this important line doublet for chromospheric plasma and magnetic field diagnostics; the CLASP2 spatial resolution is less than 2\arcsec\ and the spectral resolution is 0.1\,\r{A}.

The CLASP sounding rocket, the precursor to CLASP2, measured the linear polarization around the hydrogen Ly-$\alpha$ line at $1216$\,\r{A}, with the radially oriented spectrograph's slit extending from 20\arcsec\ off the solar limb to 380\arcsec\ on the solar disk \citep{Kano2017}. 
CLASP found a clear center-to-limb variation (CLV) in the \emph{near wings} of the Ly-$\alpha$ $Q/I$ profile, in agreement with the theoretical predictions that included PRD effects and $J$-state interference  \citep{Belluzzi2012b}. However, the data did not show any clear CLV at the \emph{center} of the Ly-$\alpha$ $Q/I$ profile, which is also sensitive to the magnetic field of the chromosphere-corona transition region for fields between approximately 10 and 100 gauss via the Hanle effect. 
This lack of CLV in the $Q/I$ line center contrasts with the clear CLV found there in 1D semi-empirical models of the solar chromosphere \citep{JTB2011} and in a 3D model of an enhanced network region \citep{Stepan2015}. 
Recently, \cite{JTB2018} applied a statistical approach to show that the lack of CLV at the center of the Ly-$\alpha$ $Q/I$ profile can be reproduced by increasing the geometrical complexity (corrugation of the transition region surface) of the 3D atmospheric model. 
Other chromospheric lines do show line-center CLV, such as the Ca {\sc ii} K line \citep{Holzreuter2007}, but their line-center originates well below the transition region---much deeper than the center of Ly-$\alpha$.

The ultimate goal of the CLASP and CLASP2 missions is to obtain UV spectropolarimetric observations to infer the magnetic field in the upper chromosphere. 
The magnetic field modifies the spectral line polarization, and thus to extract the magnetic information it is important to understand how the polarization is modified by the presence of magnetic fields in different regions of the solar atmosphere. 
As mentioned above, recent theoretical work has led to significant advances in understanding the predicted polarization in the vicinity of the \mgii $h$ \& $k$ lines; CLASP2 has provided the first ever data with which to verify these new theoretical advances.
We present here the first resolved observations of the CLV of the chromospheric \mgii $h$ \& $k$ linear polarization and compare these to theoretical predictions of the expected CLV in a semi-empirical model of the quiet solar atmosphere. 
These insights have resulted in a better understanding of the spectral line polarization produced by the solar chromosphere, and are an important part of our efforts towards the goal of measuring the chromospheric magnetic field.

\section{CLASP2 Instrument, Flight, and Data Reduction} \label{sec:instrument} 

The details of the CLASP2 instrument can be found in \cite{Tsuzuki2020}. 
CLASP2 re-uses most of the hardware from CLASP \citep{Narukage2015}, but there are a few significant changes. 
The $27$\,cm aperture Cassegrain telescope feeds into the spectrograph section after the collected light goes through a half-wave plate \citep[for the Ly-$\alpha$ wavelength,][]{Ishikawa2013}.  
For CLASP2, the primary mirror of the telescope has been re-coated \citep{Yoshida2018} and the grating completely changed, so that with the same CLASP wave plate, the dual-bandpass spectropolarimeter measures the four Stokes parameters across the \mgii lines, while the slit-jaw imager maintains the same Ly-$\alpha$ bandpass as in CLASP. 
A 196\arcsec\ long slit sits behind a Polarization Modulation Unit 
\citep[PMU,][]{Ishikawa2015}, which continuously rotates the wave plate. 
The light that passes through the slit hits the new grating in the spectrograph section which splits the light into two channels which are fed to two cameras (SP1 and SP2). 
The individual SP camera polarization analyzers are oriented orthogonal to each other. 
The linear polarizers just before the CCD (charge coupled device) cameras were also changed, and a re-imaging optic was also introduced to accommodate the change in wavelength from Ly-$\alpha$ to \mgii $h$ \& $k$ using the same CCD cameras. 
In CLASP2, the period of the PMU was changed from 4.8~seconds per rotation to 3.2~seconds per rotation to accommodate the higher intensity of the \mgii $h$ \& $k$ lines.

\begin{figure*}[ht!]
	\plotone{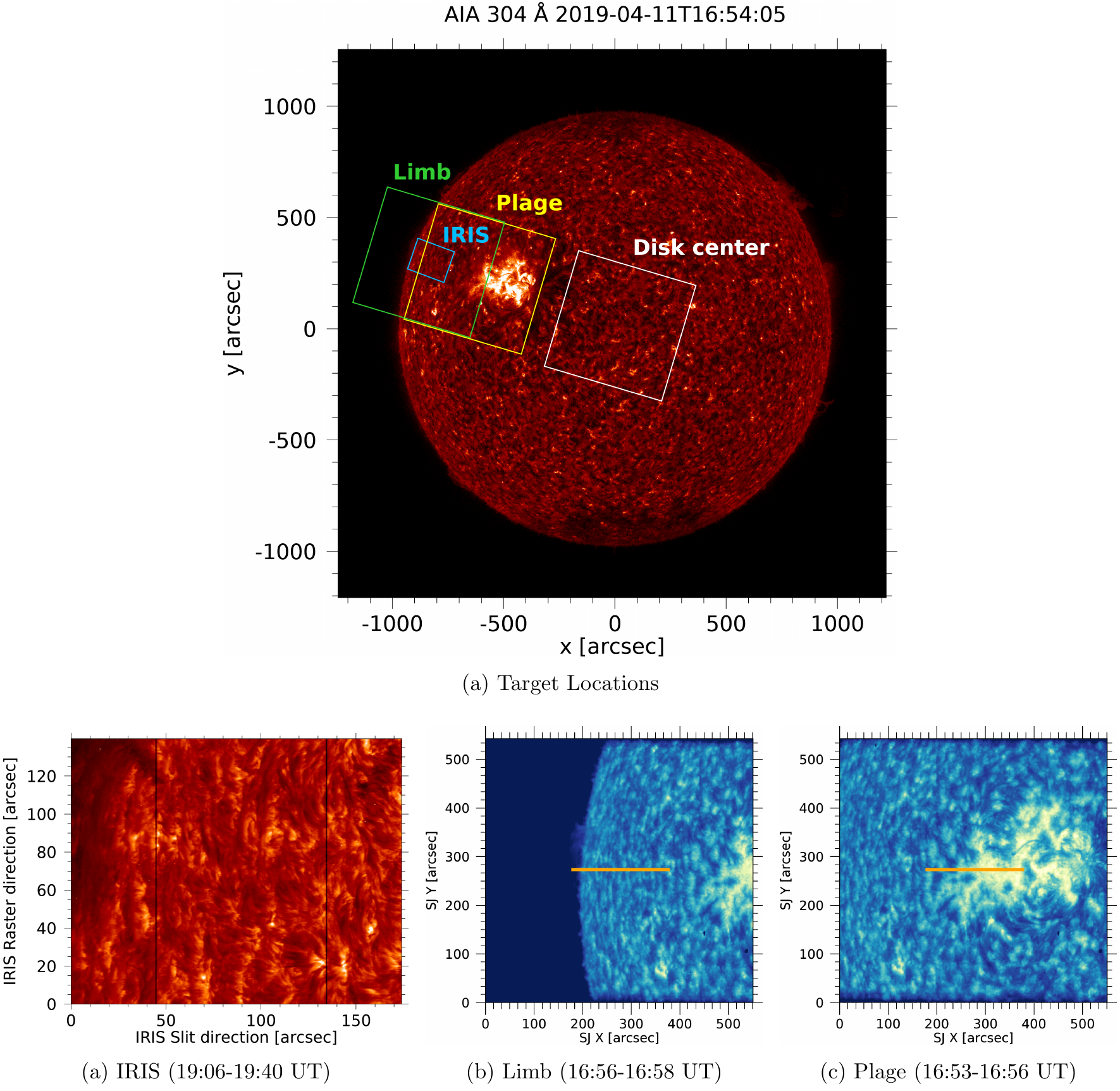}    
    \caption{(a) The CLASP2 slit-jaw FOV locations and the IRIS scan area are indicated for each target superposed on an AIA 304~\r{A}\, image. (b) IRIS raster of the limb region showing the intensity of the core of the \mgii $k$ line. The solar limb is to the left of the image. This scan was taken approximately two hours after the CLASP flight. (c) CLASP2 slit-jaw Ly-$\alpha$ image of the limb target with the slit highlighted in orange. The solar limb is to the left of the image. (d) CLASP2 slit-jaw Ly-$\alpha$ image of the plage target with the slit highlighted in orange. The solar limb is to the left of the image. } \label{fig:targets}
\end{figure*}

CLASP2 was launched from White Sands Missile Range in New Mexico on April 11, 2019 at 16:51:00\,UT and took science-quality data spanning $5$\,m $53$\,s. 
During its flight, three solar locations were targeted: disk center for calibration purposes ($15$\,sec, 16:52:52-16:53:07 UT), the edge of a strong-field plage region ($156$\,sec, 16:53:40-16:56:16 UT), and quiet sun at the limb ($141$\,sec 16:56:25-16:58:46 UT).
The top panel in Figure~\ref{fig:targets} shows these locations superposed on an SDO/AIA \citep[Solar Dynamics Observatory/Atmospheric Imaging Assembly,][]{Lemen2012} 304~\r{A} image. The slit is oriented $73.03^\circ$, $73.0^\circ$, and $72.8^\circ$ degrees counter-clock-wise from solar north in the three pointings respectively, less than $2.5^\circ$ from radial to where the slit crosses the solar limb. IRIS \citep[Interface Region Imaging Spectrograph,][]{IRISpaper} also took several co-observation raster scans. Shown in Figure~\ref{fig:targets} is a large dense 400-step limb context raster that overlays most of the CLASP2 slit (OBS 3620106078). The IRIS Raster time period is (19:06:11-19:40:37 UT), approximately two hours after the CLASP2 flight. 

This paper focuses on the spectropolarimetric (SP) slit data from the 2nd and 3rd targets, Plage and Limb, respectively. 
All of the data taken at these targets has been calibrated, demodulated, and combined into a single measurement at each target to increase the signal-to-noise ratio. 
Coordinates were cross-correlated with AIA and IRIS. 
Calibration performed to the data presented here consists of bias and dark current removal, gain correction, frame transfer smear correction, demodulation, application of the polarization response matrix for cross-talk correction (Song et al. in prep.), combining SP1 and SP2, correction of +Q to parallel to the solar limb, and time averaging.
The instrumental polarization was confirmed to be negligibly small (Song et al. in prep.) using data taken from the 1st target (Disk Center).
In this paper we only take into account the photon noise as polarization error, which is the dominant error source. 
For each plot, we indicate the 3$\sigma$ error taking into account averaging in the time, spatial, and spectral directions as appropriate for the plot.
The spectropolarimetric data presented here is a combination of  both SP cameras. Individual SP images have an exposure time of $200$\,ms and $16$ images are taken during a full PMU rotation period ($3.2$\,s). 
$47$ PMU rotations worth of data were averaged for the plage target and $43$ PMU rotations were averaged for the limb target. 
The calibrated SP data have a spectral plate scale of 0.04988 \,\r{A}/pix 
and a spatial plate scale of 0.527~\arcsec/pix.
The resolutions are 0.1\,\r{A}, and 1.2\arcsec\ respectively \citep{Yoshida2018,Song2018}.

\begin{figure*}[ht!]
\plotone{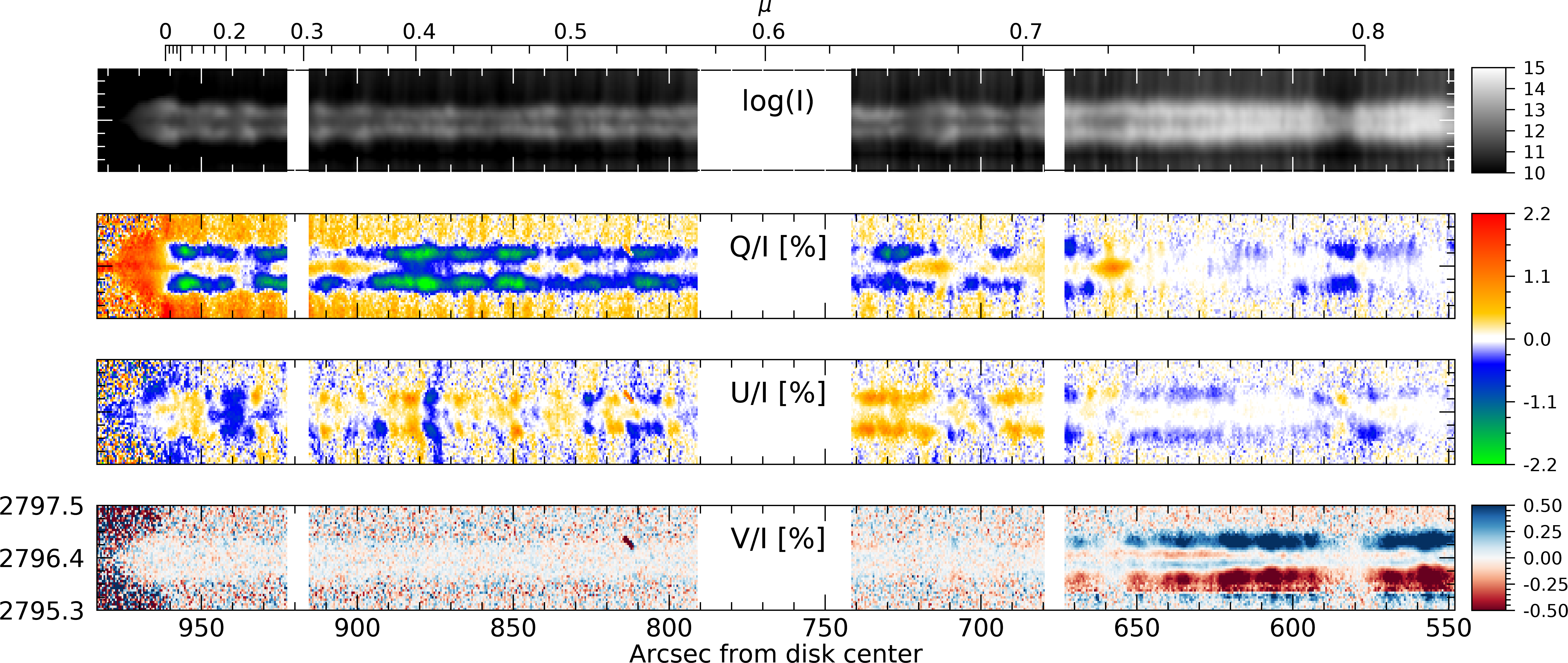}
\caption{ Stokes signals of the \mgii $k$ line for the plage (below $750$\,\arcsec, right) and the limb (above $750$\,\arcsec, left) pointings. The vacuum wavelengths are labeled in \r{A}ngstroms. The limb of the Sun is on the left side of the figure at $\mu=0$, $\mu$ is the cosine of the heliocentric angle. At the limb, the intensity of of the light outside of the line cores drops very quickly. The limb location (961.5\arcsec) in the CLASP2 wavelength range was determined by estimating where Stokes $I$ decreased by $1/e$ with respect to on-disk near-limb intensity. The reference direction for positive Stokes $Q$ is parallel to the nearest limb.
}
\label{fig:prettyspectra}   
\end{figure*}

To extend the range of the CLV beyond a single $196$\arcsec\ slit length, we combined the data from the plage and limb targets.
The slits are not exactly co-linear between those two pointings (see Figure~\ref{fig:targets}), however the $~4$\,\arcsec\ offset (perpendicular to the slit direction) is small; together the two slits cover $435.5$\arcsec\ ($0.46 R_\odot$), with a break of $49.8$\arcsec\ between the two pointings. 
All location coordinates here are presented as a distance from solar disk center. 
The slit data disk-ward of $\sim650$\arcsec\ samples a strong-field plage region, and these data are not used in the quiet sun CLV analysis that follows but are shown in the plots and figures for completeness. 
The analyses of the circular polarization data in the active region plage are presented in \cite{Ishikawa2021}.
An overview of the demodulated \mgii $k$ signal at the plage and limb targets is shown in Figure~\ref{fig:prettyspectra}, with the limb near the left edge of the plot. 
A piece of dust obscured a small portion ($\sim7$\arcsec) of the slit. 
The data under the dust has been excluded in this analysis (thin gaps at $675$\arcsec\ and $920$\arcsec\ in Figure~\ref{fig:prettyspectra}). 

\section{Observational Results}

\begin{figure}[ht!]
\plotone{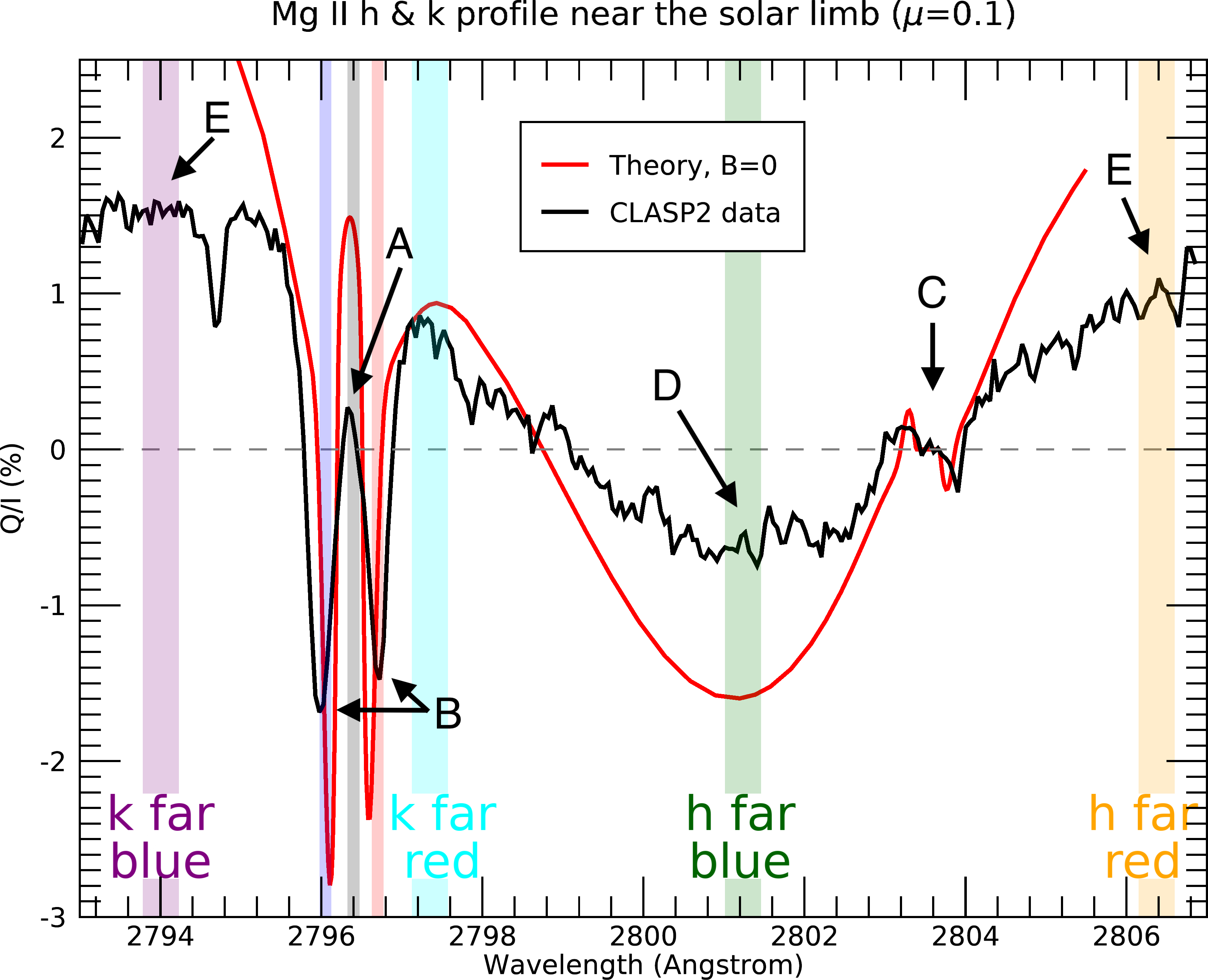}
\caption{ Comparison of theoretical (red) and observed (black) $Q/I$ line profiles near the limb. The observed profile is the average profile for the portion of the slit spanning \lr{$954.5$\,\arcsec\,$<R_\odot<958.2$\,\arcsec (7} pixels along the slit). The error bar on the black line is less than $0.1[\%]$, this error is calculated from 3$\sigma$ photon noise.
The theoretical calculation in the \textbf{B}=0 FAL-C semi-empirical model includes the effects of PRD effects and $J$-state interference \citep{Belluzzi2012}. The colors and locations of the shaded regions indicate the wavelength ranges of the data shown in Figures~\ref{fig:CLV_emissionline} and \ref{fig:CLV_outofline}. The capital letter labels correspond to descriptions in the text. The reference direction for positive Stokes $Q$ is parallel to the nearest limb.
}
\label{fig:limb_theoryVobs1}
\end{figure}

Figure~\ref{fig:limb_theoryVobs1} shows a comparison of the $Q/I$ profile observed by CLASP2 near the limb ($\mu\approx0.1$) 
with the theoretical $Q/I$ profile for the \emph{zero field case} obtained by \cite{Belluzzi2012} in the semi-empirical model C of \cite{Fontenla1993} (hereafter, FAL-C model).   
The theoretical $Q/I$ profile includes PRD effects and $J$-state interference when solving the radiative transfer problem. 
The combined action of these two physical ingredients results in significant $Q/I$ signals in the near and far  wings of the \mgii $h$ \& $k$ lines, which become stronger near the limb. As mentioned in \S1, the presence of a magnetic field and/or the symmetry breaking produced by the horizontal inhomogeneities and macroscopic velocity gradients of the solar chromospheric plasma can have an impact on the $Q/I$ and $U/I$ signals, and neither is included in the model results shown in Figure~\ref{fig:limb_theoryVobs1}. Nevertheless, several key features in the theoretical $Q/I$ profile corresponding to the unmagnetized FAL-C semi-empirical model (hereafter, the \textbf{B}=0 case) are clearly validated by the CLASP2 data (Figure~\ref{fig:limb_theoryVobs1}): 
\begin{enumerate}[label=\Alph*:,itemsep=15pt,parsep=2pt]
\itemsep0em
\item Positive measured \mgii $k$ line-center signal, which is 
subject to the Hanle effect depolarization in the presence of a magnetic field \citep{AlsinaBallester2016,delpinoaleman2016,delpinoaleman2020},
is weaker than the \textbf{B}=0 case.
\item Negative troughs are measured on either side of the \mgii $k$ line-center (the near wings), the blue wing is more negative than the red wing. The trough-peak-trough nature is caused by PRD effects and this lack of symmetry between the troughs results from the joint action of PRD and $J$-state interference \citep{Belluzzi2012}.
\item The \mgii $h$ line-center observation has zero polarization with a low-amplitude anti-symmetric signal in the near wings. The center of the $h$ line is intrinsically unpolarizable because the upper and lower levels both have total angular momentum $J=1/2$. The anti-symmetric signal is caused by the joint action of PRD and $J$-state interference \citep{Belluzzi2012}.
\item A negative trough observed between the \mgii $h$ \& $k$ centers is also produced by the combined action of PRD and $J$-state interference in the \textbf{B}=0 case \citep{Belluzzi2012}. The addition of a magnetic field, as in the solar case, would lead to a weaker $Q/I$ (and give rise to a $U/I$ signal) in this wavelength range due to magneto-optical effects \citep{AlsinaBallester2016,delpinoaleman2016,delpinoaleman2020}.
\item Large positive broadband polarization signals are present on both far edges of the wavelength range, again caused by PRD and $J$-state interference \citep{Belluzzi2012}. Most semi-empirical models predict larger $Q/I$ than was observed by CLASP2 \citep[see Figure 6 in][]{JTB2017}. 
\end{enumerate}

The magnetic sensitivity of the $Q/I$ signal at the center and wings of the lines is dominated by different effects, and the height of the $\tau=1$ surface is wavelength dependent. 
In the far wings, the \mgii radiation samples plasma in the upper photosphere, at an altitude of a few hundred km above the continuum $\tau=1$ surface \citep[see Figure 1 of][]{Belluzzi2012}.
In the line-center of both $h$ \& $k$, the light comes from the upper chromosphere. 
This wide range of formation heights implies that the polarization across the 
\mgii $h$ \& $k$ lines is sensitive to solar magnetic fields from the upper photosphere  up to close to the chromosphere-corona transition region \citep[see also][]{delpinoaleman2020}.  
The challenging task of determining the full stratification of the magnetic field from the intensity and polarization observed across the \mgii resonance lines is out of the scope of this paper. 
The comparisons of the observed results to the theoretically predicted results for the FAL-C model nonetheless prove important theoretical predictions about the nature of the polarization and the solar chromosphere, and it suggests that the inclusion of magnetic fields in the FAL-C model is needed to explain the lower polarization amplitudes observed by CLASP2.

\begin{figure*}[t!]
\plotone{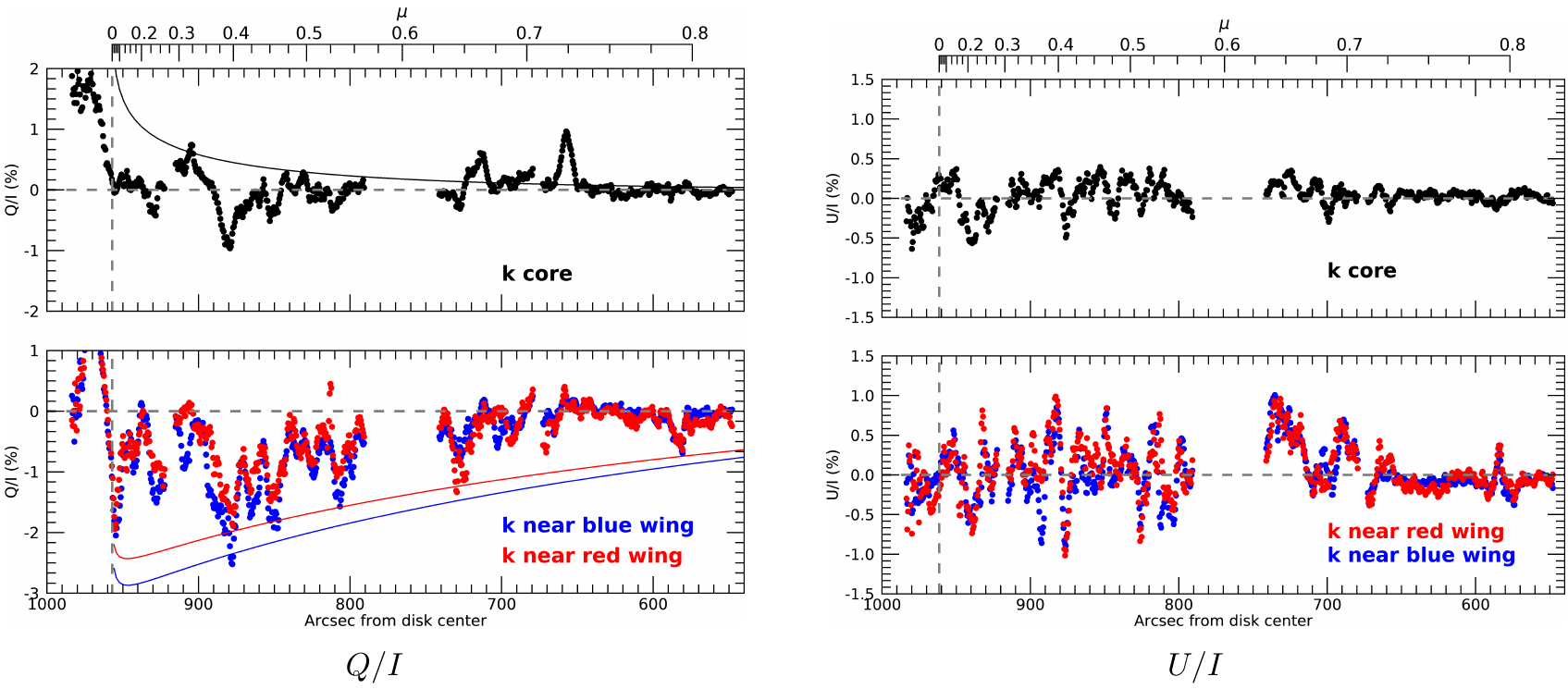}
    \caption{Center-to-limb variation of the core and near wings of \mgii $k$ in $Q/I$ and $U/I$ as a function of arcsec from disk center and of solar $\mu$. The colored dots represent the CLASP2 data and the solid 
    curves the theoretical $Q/I$ signals calculated in the unmagnetized FAL-C semi-empirical model. 
    Note that $U/I=0$ in any one-dimensional static model like FAL-C, when in the absence of inclined magnetic fields.    
    Dashed lines indicate the location of the limb and zero-polarization. The colors of the dots correspond to the shaded wavelength ranges in Figure~\ref{fig:limb_theoryVobs1}. The reference direction for positive Stokes $Q$ is parallel to the nearest limb. The 3$\sigma$ photon noise error on the dots is roughly $0.02\%$ on disk and $0.1\%$ off-limb.
    }
    \label{fig:CLV_emissionline}
\end{figure*}

The linear polarization signal at the \mgii $k$ line-center is modified by the Hanle effect in the presence of upper chromospheric magnetic fields with strengths of approximately 5 to 50 gauss \citep{Belluzzi2012,AlsinaBallester2016,delpinoaleman2016,delpinoaleman2020}. 
Unlike the very clear CLV of the $Q/I$ profile at the $k$ line-center calculated in 1D semi-empirical models of the solar atmosphere (solid curve in the top left panel of Fig.~\ref{fig:CLV_emissionline}), CLASP2 shows a signal that has strong variations along the slit (dots in Fig.~\ref{fig:CLV_emissionline}). 
Given that the linear polarization at the $k$ line center is modified by magnetic fields via the Hanle effect, this lack of a clear CLV may indicate that a fluctuating \textbf{B} in the upper chromosphere significantly modifies the CLV predicted by a simple plane parallel model atmosphere.
The spatial scale of the fluctuation of Stokes $I$ along the slit is on the order of 10\arcsec, consistent with network/internetwork spacing which would correspond with field changes. 
However, the solar chromosphere is highly inhomogeneous and dynamic; the magnetic field is not the only cause of symmetry breaking capable of modifying the scattering line polarization.
CLASP also did not observe any clear CLV at Ly-$\alpha$ line center, which was shown to be consistent with 3D model atmospheres having a highly corrugated transition region surface \citep{JTB2018}.

The \mgii $k$ blue and red near wings of the $Q/I$ profile  
show a clear CLV between about $690$\arcsec\ and $890$\arcsec, though this trend does not appear to continue to the limb (lower left panel of Fig.~\ref{fig:CLV_emissionline}). 
Via the magneto-optical effects, these near wing signals are sensitive to the depolarizing effect caused by the presence of magnetic fields in the lower solar chromosphere, with strengths similar to those that produce the Hanle effect at the $k$ line center. 
This explanation would be consistent with the observations.
The theoretical CLV calculated in the unmagnetized FAL-C semi-empirical model of the quiet sun (the solid curves in the same figure panel) generally have a stronger polarization signature than the CLASP2 measurements.

The center of the neighboring \mgii $h$ line is intrinsically unpolarizable and the distinctive antisymmetric $Q/I$ signal observed around the line center is only obtained in theoretical calculations that include both PRD effects and $J$-state interference \citep{Belluzzi2012}.  

As expected, all $U/I$ signals of the \mgii $k$ line (both, at the center and at the near wings) fluctuate around zero, without any CLV (right panels of Fig.~\ref{fig:CLV_emissionline}). 
The corresponding $Q/I$ signals also fluctuate along the spatial direction of the radially oriented slit (left panels), but the wing signals do not fluctuate around zero.    
These $Q/I$ and $U/I$ spatial fluctuations are likely caused by the local symmetry breaking due to the three-dimensional nature of the solar chromosphere, but the amplitudes of the fluctuations themselves appear to be sensitive to the magnetic field strength---note that these polarization amplitudes are damped inside the plage region disk-ward of $650$\arcsec, where the magnetic field is stronger. 
As mentioned above, the Hanle effect (which operates at the $k$ line center) and the magneto-optical effects (which operate in the near and far wings all across the \mgii doublet) tend to significantly reduce the scattering polarization amplitudes. 
This occurs mainly for magnetic fields with strengths at or above the critical field ($B_{\rm H}=22$ G) for the onset of the Hanle effect at the $k$ line center \citep{AlsinaBallester2016,delpinoaleman2016,delpinoaleman2020}.      

\begin{figure*}[t!]
\plotone{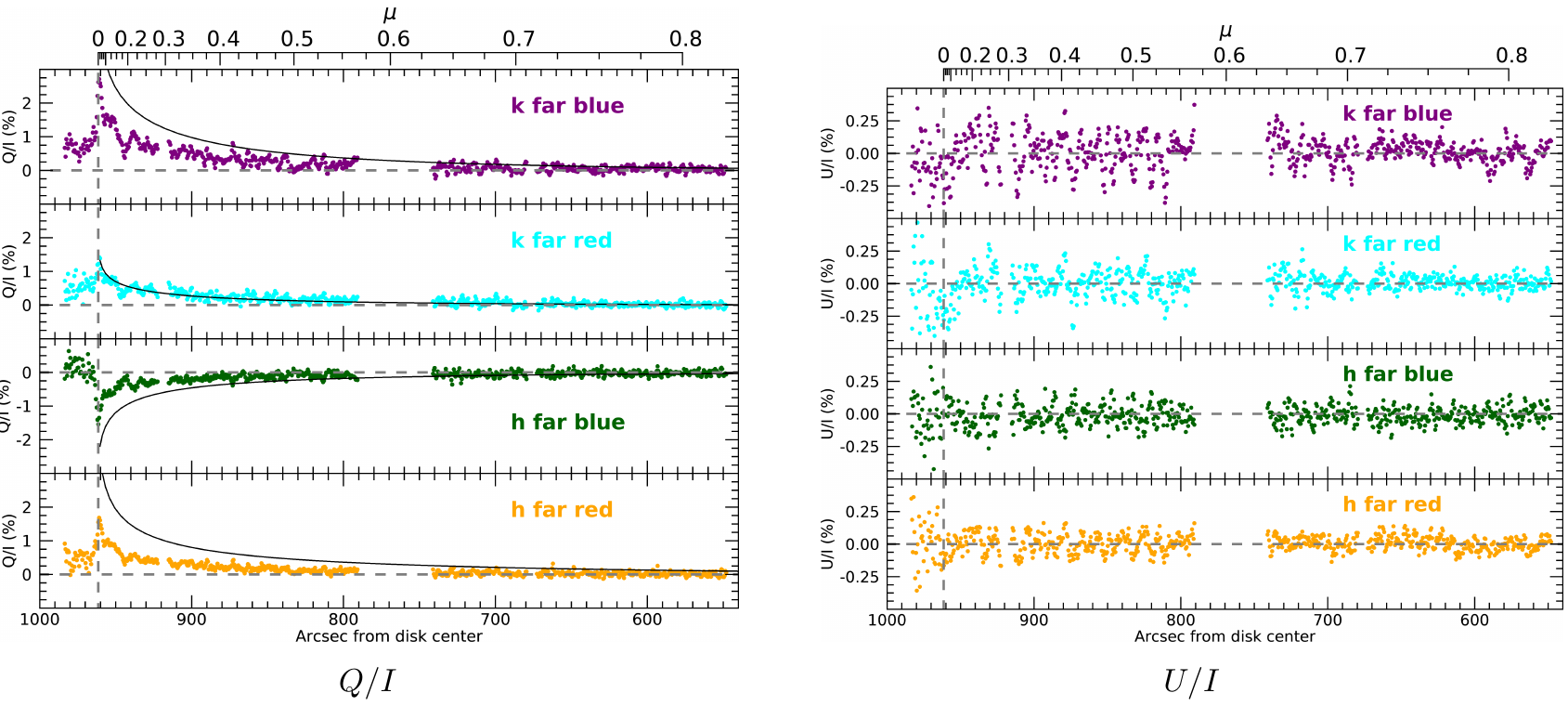}
    \caption{Center-to-limb variation of the far wings of \mgii $h$ \& $k$ in $Q/I$ and $U/I$ as a function of arcseconds from disk center and solar $\mu$, with the colored dots showing the CLASP2 data and the solid 
    curves the theoretical $Q/I$ signals calculated in the unmagnetized FAL-C semi-empirical model. 
    Note that $U/I=0$ in any one-dimensional static model like FAL-C, when in the absence of inclined magnetic fields.
    Dashed lines indicate the location of the limb and zero-polarization. The colors of the dots correspond to the shaded wavelength ranges in Fig.~\ref{fig:limb_theoryVobs1}. The reference direction for positive Stokes $Q$ is parallel to the nearest limb.
    The 3$\sigma$ photon noise error on the measured $Q/I$ is roughly $0.01\%$ on disk and $0.2\%$ off limb. The 3$\sigma$ photon noise error on the measured $U/I$ is roughly $0.05\%$ on disk and $0.25\%$ off limb.
    }
    \label{fig:CLV_outofline}
\end{figure*}

Although this was not explicitly included in the theoretical paper by \cite{Belluzzi2012}, another result from their calculations with the FAL-C semi-empirical model is that the far wings of the lines have a clear CLV in $Q/I$; the solid black curves in the left panels of Fig.~\ref{fig:CLV_outofline} show the CLV at the four far-wing wavelength bands.
The points in the same panels show the CLV observed by CLASP2 at these wavelengths. 
Interestingly, while the observed and calculated CLV match in the $k$ far red wavelength, the amplitudes of the observed CLV in the other far wing wavelengths are smaller than the theoretical ones. 
As shown by \cite{delpinoaleman2016,delpinoaleman2020} magneto-optical effects introduce magnetic sensitivity in the far wings of the \mgii $h$ \& $k$ lines, depolarizing the $Q/I$ wing signals for magnetic fields as weak as 22 G. 
These far $Q/I$ wing signals originate in the upper photosphere, and we interpret the smaller amplitudes of the observed CLV curves as due to magneto-optical depolarization by the magnetic fields of the upper solar photosphere. 

The $U/I$ far wing signals fluctuate around zero (right panels of Fig.~\ref{fig:CLV_outofline}), which is mostly as expected. 
As stated earlier, theoretical models predict that when a magnetic field is present, the magneto-optical effects can transfer signal from Stokes $Q$ to Stokes $U$ in the far wings \citep{AlsinaBallester2016, delpinoaleman2016, delpinoaleman2020}. 
The effect increases when the longitudinal component of the magnetic field is stronger and when $Q$ is greater, such as very close to the limb where the $Q/I$ wing signal rises according to its CLV. The amplitude of the $U/I$ signals very near the limb ($0<\mu<0.2$) in the far wings may thus show a CLV due to these magneto-optical effects (Fig.~\ref{fig:CLV_outofline}, right), but the observations are not conclusive. 
More work is in progress to explore this possibility.

\section{Discussion} \label{sec:discussion}
We have presented the first ever 
spectrally resolved measurements of the linear polarization across the \mgii $h$ \& $k$ lines of the solar chromosphere, with a spatial resolution of less than 2\arcsec.
These unprecedented measurements provide unique data to confirm or refute our current theoretical understanding of the physical mechanisms that control the polarization of these strong ultraviolet lines, which encode important information on the magnetism of the solar chromosphere.  
The focus of this paper is the linear polarization observed by CLASP2 in the quiet sun, with emphasis on the center-to-limb variation (CLV).   

The key findings are summarized as follows:
\begin{enumerate}[label=(\arabic*),itemsep=10pt,parsep=0em]
\item The overall shape of the $Q/I$ polarization signal is consistent with theoretical predictions  
when PRD and $J$-state interference are included in the radiative transfer calculations \citep{Belluzzi2012}, proving the importance of both of these effects for the \mgii resonance lines (Fig.~\ref{fig:limb_theoryVobs1}).
\item Although no clear CLV is seen at the $Q/I$ line center of \mgii $k$ (Fig.~\ref{fig:CLV_emissionline}, upper left panel), the observed line-center signal does not fluctuate around the zero polarization level as clearly as CLASP measured for the Ly-$\alpha$ $Q/I$ line-center signal \citep{Kano2017}. The center of the hydrogen Ly-$\alpha$ line originates very close to the corrugated surface that delineates the chromosphere-corona transition region, and the observational fact that the CLV of its $Q/I$ line center signal fluctuates around zero could be explained in terms of the geometrical complexity of such transition region layer \citep{JTB2018}. 
The \mgii $k$ line center $Q/I$ patterns observed by CLASP2 could indicate that both magnetic field variations and corrugation of the originating surface contribute to the observation.  
\item CLV is found in the near wings of \mgii $k$ in $Q/I$ (Fig.~\ref{fig:CLV_emissionline}, lower left panel).
The CLV is clearly inferred disk-ward of $\mu=0.4$, but there is a break in the pattern closer to the limb.
Depolarization by magneto-optical effects produced by chromospheric magnetic fields could be responsible for the disparity between the model and observations, but more theoretical work is needed to be able to model the linear polarization in the wings of strong resonance lines taking into account the combined action of magneto-optical effects and symmetry breaking caused by horizontal atmospheric inhomogeneities.
\item The predicted antisymmetric $Q/I$ signal around the \mgii $h$ line center was validated (Fig.~\ref{fig:limb_theoryVobs1}, C). This very distinctive signal is only seen in the theoretical models when both PRD and $J$-state interference are included \citep{Belluzzi2012}. 
\item The broadband $Q/I$ CLV trends in the far wings of \mgii $h$ \& $k$ (Fig.~\ref{fig:CLV_outofline}, left panels) are consistent with theoretical calculations when PRD effects, $J$-state interference, and magneto-optical effects are accounted for \citep{delpinoaleman2016, delpinoaleman2020, AlsinaBallester2016}. 
\end{enumerate}

More work is needed to fully understand these new observations. 
Nevertheless, the comparison of the unique data provided by CLASP2 with the theoretical results from the 1D radiative transfer investigations carried out over the last few years \citep{Belluzzi2012, AlsinaBallester2016, delpinoaleman2016, delpinoaleman2020} clearly show that we have a good physical understanding of the mechanisms that control the linear and circular polarization across the \mgii $h$ \& $k$ lines. 

The next challenge on the theoretical front will be to solve the 3D radiative transfer problem taking into account the PRD effects and $J$-state interference. 
Regarding observations, it is clear that routine spectropolarimetric observations of this sort would result in rich scientific information about the magnetic field that would benefit to a broad solar community, and so we strongly advocate for a solar space telescope equipped with a CLASP-like spectropolarimeter for probing the upper solar chromosphere.

\acknowledgments

CLASP2 is an international partnership between NASA/MSFC, NAOJ, JAXA, IAC, and IAS; additional partners include ASCR, IRSOL, LMSAL, and the University of Oslo. The CLASP2 team acknowledges S. Ishikawa, who led the development of the critical component of the polarization modulation unit (PMU) at ISAS/JAXA.
The Japanese participation was funded by ISAS/JAXA as a Small Mission-of-Opportunity Program, JSPS KAKENHI Grant numbers JP25220703 and JP16H03963, 2015 ISAS Grant for Promoting International Mission Collaboration, and by 2016 NAOJ Grant for Development Collaboration. 
The USA participation was funded by NASA Award 16-HTIDS16\_2-0027. 
The Spanish participation was funded by the European Research Council (ERC) under the European Union's Horizon 2020 research and innovation programme (Advanced Grant agreement No. 742265). 
The French hardware participation was funded by CNES funds CLASP2-13616A and 13617A.
The Swiss participation was funded by SNSF grants 200021-175997 and CRSII5-180238. 
B.D.P. was supported by NASA Contract NNG09FA40C (IRIS).

%

\vspace{5mm}






\bibliography{V5_CLASP2_CLV_apj}{}
\bibliographystyle{aasjournal}



\listofchanges
\end{document}